\documentclass[11pt,a4paper]{article}
\usepackage{jheppub}
\pdfoutput=1
\usepackage[utf8]{inputenc}
\usepackage{amsfonts,amsmath,amssymb}
\usepackage{graphicx}
\usepackage{multirow}
\usepackage{verbatim}
\usepackage{bm}
\usepackage[textsize=scriptsize,backgroundcolor=red!70,linecolor=red]{todonotes}
\usepackage{paralist}
\usepackage{subfig}
\usepackage[normalem]{ulem}

\newcommand{\ii}{\mathrm{i}}
\newcommand{\ee}{\mathrm{e}}

\newcommand{\ket}[1]{\mid #1 \rangle}
\newcommand{\bra}[1]{\langle #1 \mid}
\newcommand{\abs}[1]{\big\lvert #1 \big\rvert}

\newcommand{\ie}{i.e.}
\newcommand{\eg}{e.g.}

\newcommand{\ud}{\ensuremath{\mathrm{d}}}

\newcommand{\chisq}{\ensuremath{\chi^{2}}}

\newcommand{\porpb}{\ensuremath{\overset{\scriptscriptstyle \left(-\right)}{p}}}
\newcommand{\fform}{\ensuremath{\mathcal{F}}}
\newcommand{\bvec}[1]{\ensuremath{\bm #1}}
\newcommand{\dof}{\ensuremath{\text{d.o.f}}}
\newcommand{\apr}{\ensuremath{\alpha^\prime_P}}
\renewcommand{\Im}{\ensuremath{\mathfrak{Im}}}
\renewcommand{\Re}{\ensuremath{\mathfrak{Re}}}

\title{Unitarisation dependence of diffractive scattering in light of high-energy collider data}
 
\author[a]{Arno Vanthieghem,}
\author[b]{Atri Bhattacharya,}
\author[b]{Rami Oueslati,} 
\author[b]{and Jean-René Cudell}

\emailAdd{vanthieg@slac.stanford.edu}
\emailAdd{a.bhattacharya@uliege.be}
\emailAdd{rami.oueslati@uliege.be}
\emailAdd{jr.cudell@uliege.be}

\affiliation[a]{High Energy Density Science Division (HEDS),
SLAC National Accelerator Laboratory, Menlo Park, California 94025, USA}
\affiliation[b]{Space sciences, Technologies and Astrophysics Research (STAR) Institute,
                Université de Liège, Bât.~B5a, 4000 Liège,
                Belgium}
\date{\today}

\abstract{
We study the consequences of high-energy collider data on the best fits to total,
elastic, inelastic, and single-diffractive cross sections for $pp$ and $p
\bar{p}$ scattering using different unitarisation schemes.
We find that the data are well fitted both by eikonal and U-matrix schemes, but that diffractive data
prefer the U-matrix.
Both schemes may be generalised by means of an additional parameter;
however, this yields only marginal improvements to the fits.
We provide estimates for $\rho$, the ratio of the real part to the imaginary part
of the elastic amplitude, for the different fits.
We comment on the effect of the different schemes on present and future cosmic
ray data.
}

\begin{document}
\maketitle

\section{\label{sec:intro}Introduction}

High-energy hadronic scattering may be described by
Reggeon exchanges
(see, \eg\ \cite{Donnachie:2013xia} and references therein) and for center-of-momentum
energies $\sqrt{s}$ larger than 100~GeV, the only trajectory that matters is that of the pomeron.
However, at energies of a few TeV and higher, the growth of the pomeron term
leads to violation of the black-disk limit 
\cite{Gotsman:1993ux,Shoshi:2002in,Selyugin:2004sy} and  eventually of unitarity. Unitarity can be enforced in high-energy $pp$ and $p\bar{p}$
interactions by the inclusion of multiple exchanges, which act as a cut to the elastic scattering amplitude.
Different unitarisation schemes have been discussed in the literature \cite{Cudell:2008yb}  but all of them rely on phenomenological arguments in the absence of a comprehensive quantum chromodynamics treatment. 

The effect of unitarisation on the growth of $p\porpb$ cross-sections becomes
important when considering proton-proton scattering cross sections at
the LHC where the centre-of-momentum energies extend up to 13 TeV.
Measurements of the total, elastic, inelastic, and diffractive $pp$
cross sections by the different LHC experiments --- ALICE \cite{Abelev:2012sea},
ATLAS \cite{Aad:2014dca,Aaboud:2016ijx,Aad:2011eu,Myska:2017iqc},
CMS \cite{Sirunyan:2018nqx},
LHCb \cite{Aaij:2018okq}, and
TOTEM \cite{Antchev:2011vs,Antchev:2013gaa,Antchev:2013iaa,
Antchev:2013paa,Antchev:2017dia,Antchev:2013haa} --- add to existing $p\porpb$
scattering data at lower energies from previous generation experiments
at the  S$p\bar p$S \cite{Bozzo:1984rk,Alner:1986iy} and the TeVatron
\cite{Abe:1993xx,Amos:1990jh,Amos:1992zn,Abe:1993wu,Avila:1998ej,Avila:2002bp}.
This extensive wealth of data allows us to constrain the nature of unitarisation
governing these interactions with an improved degree of accuracy.

Differences in cross sections that depend on the choice of the unitarisation
scheme are expected to show up at very high energies --- at 10 TeV
and higher --- and therefore may influence predictions for cosmic-ray collisions
with atmospheric nuclei at ultra-high energies.
Showering codes, such as \verb+SIBYLL+ \cite{Engel:2019dsg} and
\verb+QGSJET+ \cite{Ostapchenko:2019few}, used to simulate and reconstruct these events from observations
of secondaries have historically used the eikonal scheme (see
\cite{Anchordoqui:2004xb} for a review).
In the context of ongoing ultra-high energy cosmic-ray experiments, \eg\
the Pierre Auger Observatory \cite{ThePierreAuger:2015rma}, the Telescope Array
Project \cite{Kawai:2008zza}, and IceTop \cite{IceCube:2012nn}, an
investigation of the dependence of cross sections on different unitarisation schemes  
assumes paramount importance. 

In \cite{Bhattacharya:2020lac}, we examined the effect of including
up-to-date collider data for total, elastic, and inelastic cross sections.
We found nearly identical cross sections for the three irrespective of
the unitarisation scheme used.
In the current work, we focus on the effect of incorporating diffractive data into
the fits.
Diffractive scattering in $ 2 \mapsto 2 $ interactions, where either one or
both final state particles break up into jets, becomes increasingly important
as the interaction energy increases.
In these interactions, the final state(s) being no longer expressible
in terms of hadronic eigenstates,
the calculation of the corresponding scattering amplitudes
requires the invocation of a rotated eigenstate basis as described in 
the Good-Walker mechanism \cite{Good:1960ba}.

The present work is organised as follows.
In Section
\ref{sec:unit}, we briefly recapitulate the theory of unitarisation in $p\porpb$ scattering
and the different schemes that have been proposed in the literature.
In Section \ref{sec:GW} we explain the Good-Walker representation
\cite{Good:1960ba}.  In section \ref{sec:fits} we list the various parameters
defining our fits.
Additionally, we list all the $p\porpb$ scattering data that are used to determine our
best fits.
Finally, in Section \ref{sec:result} we give our results and discuss them in
light of the existing literature, drawing our conclusions.

\section{\label{sec:unit}Brief survey of unitarisation schemes and fit to non-diffractive forward data}

The differential cross section for elastic scattering  may be expressed in terms of the elastic amplitude $A(s,t)$
as
\newcommand{\bq}{\bvec{q}}
\newcommand{\bb}{\bvec{b}}
\begin{equation}
	\frac{\ud\sigma_{el}}{\ud t} = \frac{\abs{A(s,t)}^2}{16\pi s^2}\,,
	\label{eq:dsigmadt}
\end{equation}
where $t = -\bq^2$ is the square of the momentum transfer.
At low energy, the term in $A(s,t)$ 
responsible for the growth of the cross section with $s$ can be parameterised \cite{Cudell:2005sg} using the pomeron 
trajectory $\alpha(t)$, 
the proton elastic form factor $\fform_{pp}(t)$ and the coupling pomeron-proton-proton $g_{pp}$, as
\begin{equation}
	a(s,t) =g_{pp}^2\, \fform_{pp}(t)^2 \left( \frac{s}{s_0} \right)^{\alpha(t)}\, \xi(t),
	\label{eq:amp_t}
\end{equation}
with $\xi(t)$ the signature factor
\begin{equation}
	\xi(t) =-e^{-i\pi\alpha(t)\over 2} .
	\label{eq:amp_t2}
\end{equation}
We shall consider here a dipole form factor, which is close to the best functional form \citep{Cudell:2005sg},
although the exact functional form is not very important as we consider only integrated quantities in this paper:
\begin{equation}
\fform_{pp}={1\over (1-t/t_{pp})^2}
\label{form}
\end{equation}
The pomeron trajectory is close to a straight line \cite{Cudell:2003dz}, and we take it to be
\begin{equation}
	\alpha(t)=1+\epsilon+\apr t.
	\label{eq:amp_t3}
\end{equation}
At high energy, the growth of this pomeron term and eventual violation of unitarity is most
clearly seen in the impact-parameter representation, where 
the Fourier transform of the amplitude $a\left( s, t \right)$ rescaled by $2s$ is equivalent to a partial wave
\begin{equation}
	\chi(s, \bvec{b}) = \int \frac{\ud^2\bvec{q}}{\left( 2\pi \right)^2}
	\frac{a(s,t)}{2s} \ee^{\ii \bvec{q}\cdot \bvec{b}}.
	\label{eq:Gsb}
\end{equation}
The norm of this partial wave at small $\abs{\bb}$ exceeds unity around $\sqrt{s}=2$ TeV \cite{Cudell:2003dz}.

To solve this problem, one introduces unitarisation schemes which map the amplitude $\chi(s,\bb)$  to
the physical amplitude $X(s,\bb)$. The latter reduces to $\chi(s,\bb)$ for small $s$, is confined to
the unitarity circle $\abs{X(s, \bvec{b})-i}
\leqslant 1 $, and bears the same relation as Eq.(\ref{eq:Gsb}), but this time to the physical amplitude:
 \begin{equation}
	X(s, \bvec{b}) = \int \frac{\ud^2\bvec{q}}{\left( 2\pi \right)^2}
	\frac{A(s,t)}{2s} \ee^{\ii \bvec{q}\cdot \bvec{b}}.
	\label{eq:Xsb}
\end{equation}

The most common scheme is the eikonal scheme, and it has been derived for structureless bodies, in optics, in potential 
scattering and in QED. Another proposed scheme is the U matrix scheme, which can be motivated by a form of 
Bethe-Salpeter  equation \cite{Logunov:1971jy}. Probably neither of these is correct in QCD, but going from one to the other permits an 
evaluation of the systematics linked to unitarisation.

In the following, we shall actually use generalised versions of the
schemes, which include an extra parameter $\omega$ 
\cite{Cudell:2008yb}:
\begin{equation}
	X_E(s, \bvec b) = \frac{\ii}{\omega}
	                  \left[ 1 - \ee^{\ii \omega \chi(s, \bvec b)}
			              \right]\,,
	\label{eq:eikx}
\end{equation}
while the generalised U-matrix scheme requires:
\begin{equation}
	X_U(s,\bvec b) = \frac{\chi(s, \bvec b)}{1 - \ii \omega
		                               \chi(s, \bvec b)}.
	\label{eq:umatx}
\end{equation}
In both cases, the asymptotic value of $X(s\to\infty,\bb)$ is $1/\omega$, hence the traditional values
for $\omega$ are 1 for the standard eikonal \cite{Logunov:1971jy}, and
$1/2$ for the standard U matrix
\cite{Savrin:1976zn}.
Both schemes map the amplitude $\chi(s, \bvec{b})$ into the
unitarity circle for  $ \omega \geqslant 1/2 $. In terms of partial waves, the maximum inelasticity is reached for
$ X(s,\bvec{b}) = \ii $.

The total and elastic scattering cross sections may be readily expressed in these
representations as 
\begin{align}
	\sigma_{{tot}} &= 2 \int\! \ud^{2} \bvec{b}\ \mathfrak{Im} \left( X(s,\bvec{b}) \right),
  &
	\sigma_{{el}}  &= \int\! \ud^{2} \bvec{b}\ \abs{X(s, \bvec b)}^{2}.
	\label{eq:sigma}
\end{align}

Hence these unitarised schemes naturally lead to expressions for the total, elastic, and hence inelastic, cross sections.
We shall now use them to fit all the data in $ p\porpb $ scattering above 100 GeV, for which lower trajectories have a negligible effect.
This includes the following:
\begin{itemize}
  \item $pp$ total and elastic cross sections from
        TOTEM \cite{Antchev:2011vs,Antchev:2013gaa,Antchev:2013iaa,
        Antchev:2013paa,Antchev:2017dia}, and ATLAS \cite{Aad:2014dca,Aaboud:2016ijx};
  \item $p\bar{p}$ total and elastic cross sections from CDF \cite{Abe:1993xx},
        E710 \cite{Amos:1990jh,Amos:1992zn}, and E811
        \cite{Avila:1998ej,Avila:2002bp} experiments at TeVatron; and UA4 at
         S$\rm p\bar p$S \cite{Bozzo:1984rk};
  \item Direct measurements of inelastic cross sections, \ie\ not derived from
	      total and elastic measurements, from UA5 at  S$\rm p\bar p$S
	      \cite{Alner:1986iy}, ATLAS \cite{Aad:2011eu,Myska:2017iqc}, LHCb
	      \cite{Aaij:2018okq}, ALICE \cite{Abelev:2012sea}, and TOTEM
        \cite{Antchev:2013haa}.
\end{itemize}

This gives a total of 37 data points. In the next section, we shall also consider 6 extra data points:
 \begin{itemize}
 \item Single diffractive $p\bar{p}$ cross sections from UA5
        \cite{Alner:1986iy,Alner:1987wb} and E710 \cite{Amos:1992jw}; and
  \item $pp$ single diffractive cross sections at various energies measured at
        ALICE \cite{Abelev:2012sea}. 
\end{itemize}
The resulting fit leads to the following parameters of Table \ref{tab:el}.
\begin{table}
 \begin{center}
\begin{tabular}{|c||c|c|c|c|c|}
   \hline
   Scheme   & $\epsilon $
            & $\apr $ (GeV$^{-2}$)
            & $ g_{pp} $
            & $ t_{pp}$ (GeV$^2$)
            & $ \chisq/\dof $\\
            \hline
  U-matrix
            & $ 0.10 \pm 0.01 $
            & $ 0.37 \pm 0.28 $
            & $ 7.5  \pm 0.8 $
            & $ 2.5 \pm 0.6 $
            & $ 1.436 $ \\  
   Eikonal
            & $ 0.11 \pm 0.01 $
            & $ 0.31 \pm 0.19 $
            & $ 7.3  \pm 0.9 $
            & $ 1.9  \pm 0.4$
            & $ 1.442 $ \\
   \hline
\end{tabular}

  \end{center}
  \caption{\label{tab:el}\chisq/\dof\ and best-fit parameters obtained using the
           eikonal ($ \omega = 1 $) and U-matrix ($ \omega^\prime = 1/2 $)
           unitarisation schemes without diffractive
           data \cite{Bhattacharya:2020lac} .
          }
\end{table}
\paragraph{}

\section{\label{sec:GW}Unitarisation and diffraction}
The implementation of diffraction within a unitarisation scheme at high energy has to confront two questions: how does one describe the diffractive amplitude at the Born level, and how does one embed that amplitude 
within a unitarisation scheme? 

The first questions has two answers. On the one hand, the asymptotic answer is that , for high-mass final states, one should use the triple-reggeon vertices. However,
as the masses considered are not necessarily large, one must consider a variety of reggeons lying
on trajectories below that of the pomeron\cite{Donnachie:2003ec}, and to include not only subdominant trajectories (with intercept of the order of $1/2$)  but also sub-subdominant ones (with an intercept of the order of 0). This introduces a multitude of parameters,
of the order of the number of high-energy data points available.  

On the other hand, it is possible to consider a generic diffractive state $D$ and
the vertex $p+I\!\!P\to D$. A priori, this implies the consideration of a large number of channels for the diffractive state $D$, and the introduction of many parameters.
However, it has been shown in \cite{Gotsman:1999xq} that 
for inclusive cross section, the consideration of one generic diffractive state $\ket{\Psi_D} $ is sufficient, and that 
adding other states does not significantly improve the description of the data.
One however looses the information about the mass of the diffractive state.

We will concentrate in
this paper on inclusive quantities,  and on a generic diffractive $D$, which
is the seed of high-energy pions, and hence of high-energy muons, in cosmic ray showers. This makes them  
of particular interest in view of the muon anomaly at ultra-high energies (see \cite{Aab:2021zfr} and 
references therein).

The second question concerns the description of multiple exchanges, which are expected to be important at 
ultra-high energies. The problem is to include insertions that contain the $p I\!\!P D$, the $D I\!\!P p$ and 
$D I\!\!P D$ vertices, and re-sum them. Solutions to this problem have been proposed by Gotsman, Levin, and Maor (GLM) \cite{Gotsman:1999xq,Gotsman:2014pwa} and further explored by Khoze, Martin, and Ryskin \cite{Khoze_2013, Khoze:2018kna,Khoze_2018b}  using the Good-Walker model \cite{Good:1960ba}. We shall
adapt their method, originally proposed for the eikonal unitarisation, to any scheme, and more specifically to the $U$-matrix unitarisation scheme.

At the Born level, the interaction of a proton with a pomeron can leave the proton intact or turn it into a diffractive state $D$. GLM argue that it is possible to define two states $\ket{\Psi_1} $ and $\ket{\Psi_2} $ which are not modified
by the interaction with a pomeron:
\begin{subequations}
\begin{align}
  \ket{\Psi_p} &= \cos\theta \ket{\Psi_1} + \sin\theta \ket{\Psi_2}\, \text{, and } \\
  \ket{\Psi_{D}} &= -\sin\theta \ket{\Psi_1} + \cos\theta \ket{\Psi_2}\,,
\end{align}
\label{eq:GWstates}
\end{subequations}
with $\theta$ an arbitrary angle. In this representation, the final states for elastic, single diffractive, and
double diffractive amplitudes are given by $ \ket{ \Psi_p\Psi_p } $, $
\ket{\Psi_p\Psi_D} $, and $\ket{\Psi_D\Psi_D}$ respectively.

Before we unitarise, we need the Born-level amplitudes $a_{ij}(s,t)= \bra{\Psi_i\Psi_j} \hat{T} \ket{\Psi_i \Psi_j}$, for $i,j=1,2$. We shall assume that the pomeron is a simple pole at the Born level, so that the amplitudes can be factorised in $t$ space as e.g.
\begin{eqnarray}
a_{pp\to pp}&=&\langle pp|T|pp\rangle=V_{pp}(t)^2 R(s,t)\\
a_{pp\to pD}&=&a_{pp\to Dp}=\langle pp|T|pD\rangle=V_{pp}(t) V_{pD}(t) R(s,t)\\
a_{DD\to DD}&=&\langle DD|T|DD\rangle= V_{DD}(t)^2 R(s,t)
\end{eqnarray}
with $R(s,t)=\left( \frac{s}{s_0} \right)^{\alpha(t)}
                   \xi(t)$, and $V_{ab}(t)$ the vertex functions.
All processes can be described using 3 functions, $V_{pp}$, $V_{DD}$ and $V_{pD}=V_{Dp}$.  We take them as
\begin{equation}
V_{ab}=g_{ab} \fform_{ab}(t)
\label{vertices}
\end{equation}
where $a$ and $b$ are either $p$ or $D$, $g_{ab}$ are the coupling strengths and  $\fform_{ab}(t)$ is a form factor, with $\fform_{ab}(0)=1$.
The nature of form factors for the eigenstates $\Psi_{\{1,2\}}$ cannot be
determined from experiments, therefore we invert the relations in Eq.~\eqref{eq:GWstates} to express $\Psi_{\{1,2\}}$ in terms of $\Psi_{\{p,D\}}$.
This allows us to work with the proton and diffractive state form factors; we assume the form factor for the latter is similar to that of the proton.
The two GLM states 
\begin{eqnarray}
|\Psi_1\rangle&=&\cos\theta|p\rangle-\sin\theta|D\rangle\\
|\Psi_2\rangle&=&\sin\theta|p\rangle+\cos\theta|D\rangle,
\end{eqnarray}
correspond to amplitudes
\begin{equation}
a_{ij\to kl}=\langle ij|T|kl\rangle=V_{ik}(t)V_{jl}(t) R(s,t),\ i,j=1,2
\label{chiij}
\end{equation}
which will be purely elastic if $V_{12}=V_{21}=0$. This leads to 
\begin{equation}
\tan(2\theta(t))={ 2V_{pD}(t)\over V_{DD} (t)- V_{pp}(t)}
\label{tangent}
\end{equation}
and
\begin{eqnarray}
	V_{11}(t) &=& V_{pp} (t)\cos^2(\theta) +V_{DD}(t)\sin^2(\theta) 
	-V_{pD}(t)\sin(2\theta)\\ 
	V_{22} (t)&=& V_{pp}(t) \sin^2(\theta) +V_{DD}(t)\cos^2(\theta) 
	+V_{pD}(t)\sin(2\theta).
	\label{GLMBorn}
\end{eqnarray}
Hence at this point, we have traded three amplitudes $V_{ab}$ for two amplitudes $V_{ii}$ and an angle.
We do not know, at the born level, how any of these should behave, except for $V_{pp}(t)$, for which the
parameterisation (\ref{vertices}) is a good representation at low energy \citep{Cudell:2005sg}. One can assume the same functional form holds for $V_{DD}$, hence we keep these two parameterisations. Following GLM \cite{Gotsman:1993ux}, we choose $\theta$ as a final input.
Clearly, it depends on $t$. However, as we shall be considering
integrated cross sections, and as the $t$ dependencies of the various $V_{ab}$ are not expected to be very different,  it is reasonable to approximate
\begin{equation}
\tan(2\theta(t))\approx \tan(2\theta(0))={ g_{pD}+g_{D p}\over g_{DD} - g_{pp}}
\end{equation}
and keep it as a parameter.
To translate this into a specific expression for $V_{11}$ and $V_{22}$, we eliminate $V_{pD}$ using Eq.~(\ref{tangent}).
This leads to
\begin{eqnarray}
	V_{11}(t) &=& {\cos^2(\theta) V_{pp} (t)-\sin^2(\theta) V_{DD}(t)\over \cos(2\theta)}\\
	V_{22}(t) &=& {\cos^2(\theta) V_{DD} (t)-\sin^2(\theta) V_{pp}(t)\over \cos(2\theta)}
	\label{VBOCBorn}
\end{eqnarray}

These can be used to build the amplitudes that will enter into the unitarisation schemes, using Eq. (\ref{chiij}).
One can thus obtain the elastic, single-diffractive and double-diffractive amplitudes from three purely elastic amplitudes \cite{Gotsman:1999xq,Gotsman:2014pwa}, given the fact that $a_{12\to 12}=a_{21\to 21}$:
\begin{subequations}
\begin{align}
a_{pp\to pp}  &=  \cos^4(\theta) a_{11\to11} + 2 \cos^2(\theta)\sin^2(\theta)  a_{12\to 12}  + \sin^4(\theta) a_{22\to 22} \\
a_{pp\to pD} &= 
                \cos(\theta) \sin(\theta)  \nonumber\\  &\times( - \cos^2(\theta) a_{11\to 11} + (\cos^2(\theta) - \sin^2(\theta)) a_{12\to 12} + \sin^2(\theta) a_{22\to 22} ) \\
a_{pp\to DD}  &=  \cos^2(\theta)  \sin^2(\theta)  ( a_{11\to 11} - 2a_{12\to 12}  + a_{22\to 22} ).
\end{align}
\label{eq:amp}
\end{subequations}

At this point, it is easy to unitarise the amplitudes $a_{ij\to ij}(s,t)$, following what was done in Section 2 for elastic scattering. One goes into impact parameter space to obtain the corresponding $\chi_{ij\to ij}(s,{\bf b})$, replaces the amplitudes at the Born level by 
their unitarised version, as Eqs. (\ref{eq:eikx}) and (\ref{eq:umatx}):
\begin{subequations}
\begin{align}
	X^{(E)}_{ij\to ij}(s, \bvec b) &= \frac{\ii}{\omega}
	                  \left[ 1 - \ee^{\ii \omega \chi_{ij\to ij}(s, \bvec b)}
			              \right]\,\\
	X^{(U)}_{ij\to ij}(s,\bvec b) &= \frac{\chi_{ij\to ij}(s, \bvec b)}{1 - \ii \omega
		                               \chi_{ij\to ij}(s, \bvec b)}.
	\label{eq:unita}
	\end{align}
\end{subequations}
and obtains the amplitudes of interest as in Eq.~(\ref{eq:amp}):
\begin{subequations}
\begin{align}
X_{el}  &=  \cos^4(\theta) X_{11\to11} + 2 \cos^2(\theta)\sin^2(\theta)  X_{12\to 12}  + \sin^4(\theta) X_{22\to 22} \\
X_{sd} &= 
                \cos(\theta) \sin(\theta)  \nonumber\\  &\times( - \cos^2(\theta) X_{11\to 11} + (\cos^2(\theta) - \sin^2(\theta)) X_{12\to 12} + \sin^2(\theta) X_{22\to 22} ) \\
X_{dd}  &=  \cos^2(\theta)  \sin^2(\theta)  ( X_{11\to 11} - 2X_{12\to 12}  + X_{22\to 22} ).
\end{align}
\label{eq:ampunit}
\end{subequations}

The relevant $ 2 \to 2 $ cross sections are then given by
\begin{subequations}
\begin{align}
  \sigma_{tot}  &= 2 \int\! \ud^2 b\ \Im\left\{ X_{el} \right\}\,;
  &\sigma_{el}  &= \int\! \ud^2 b\ \abs{X_{el}}^2\,; \\
  \sigma_{sd}   &= 2\int\! \ud^2 b\ \left(\abs{X_{sd}}^2 \right)\,;
  &\sigma_{dd}  &= \int\! \ud^2 b\ \abs{X_{dd}}^2\,;\\
  \intertext{and the $\rho$ parameter is defined by}
  \rho(s, t=0) &= \frac{\Re\left\{ X_{el}(s, t=0) \right\}}
                       {\Im \left\{ X_{el} (s, t=0) \right\}}\,.
\end{align}
\label{eq:sigmas}
\end{subequations}

\section{\label{sec:fits}Fit parameters and data}
Section 3 has introduced the basic ingredients and parameters of our model. First of all, one has of course the 
parameters of Section 2, i.e. $\epsilon$ and $\alpha'$, linked to the Pomeron trajectory $R(s,t)$, as well as
 $g_{pp}$ and $t_{pp}$, linked to the $p I\!\!P p$ vertex $V_{pp}(t)$. To describe diffractive scattering in our 
 scheme, one needs three more parameters: the $D I\!\!P D$ coupling $g_{DD}$, the scale $t_{DD}$ in the form 
 factor
\begin{equation}
\fform_{DD}(t)={1\over (1-t/t_{DD})^2}
\end{equation}
and the mixing angle $\theta$.
Finally, one can introduce the parameters $\omega$ and $\omega'$ corresponding to extended unitarisation schemes. 

Several remarks are in order at this point. First of all, we have considered the minimal GLM scheme, where we 
mix the proton with one diffractive state. This corresponds to a 2-channel unitarisation scheme. In principle, one 
could consider an $N-$channel scheme, at the cost of multiplying the number of parameters $N(N+1)/2+8$.
Given the paucity of diffractive data at high energy, going beyond $N=2$ is not possible. Note that GLM 
considered the case $N-3$, and found that there is no significant improvement \citep{Gotsman:1999xq}.

We can further limit the number of parameters by considering the two standard unitarisation schemes, i.e. fix $
\omega=1$ and $\omega'=1/2$. We have checked that varying these parameters lead to an improvement
of only $0.01$ in the $\chi^2/\dof$.

Nevertheless, even in the 2-channel scheme, one still has an over-parameterisation. The main problem comes
from the fact that there is a strong correlation between the parameters of $V_{pp}$ and those of $V_{DD}$, so
that error bars are huge. As the $pp$ parameters are determined by the fits of Section 2, we fix their values to 
their central values in that fit: $g_{pp}=7.5~(7.3)$ and $t_{pp}=2.6~(1.9)$ GeV$^2$ in the U-matrix (eikonal) 
schemes.

Since our focus is on high energy effects induced in $ p\porpb $ cross sections,
we use experimental data above 100 GeV. Together with the data set provided by the Particle Data Group~\cite{Tanabashi:2018oca}, Table~\ref{tab:dataset} includes the data from the following experiments:
\begin{itemize}
  \item $pp$ total and elastic cross sections from
        TOTEM \cite{Antchev:2011vs,Antchev:2013gaa,Antchev:2013iaa,
        Antchev:2013paa,Antchev:2017dia}, and ATLAS \cite{Aad:2014dca,Aaboud:2016ijx};
  \item $p\bar{p}$ total and elastic cross sections from CDF \cite{Abe:1993xx},
        E710 \cite{Amos:1990jh,Amos:1992zn}, and E811
        \cite{Avila:1998ej,Avila:2002bp} experiments at the TeVatron; and UA4 at the
        S$\rm p\bar p$S \cite{Bozzo:1984rk};
  \item Direct measurements of inelastic cross sections, \ie\ not derived from
	      total and elastic measurements, from UA5 at the S$\rm p\bar p$S
	      \cite{Alner:1986iy}, ATLAS \cite{Aad:2011eu,Myska:2017iqc}, LHCb
	      \cite{Aaij:2018okq}, ALICE \cite{Abelev:2012sea}, and TOTEM
        \cite{Antchev:2013haa};
  \item Single diffractive $p\bar{p}$ cross sections from UA5
        \cite{Alner:1986iy,Alner:1987wb} and E710 \cite{Amos:1992jw}; and
  \item $pp$ single diffractive cross sections at various energies measured at
        ALICE \cite{Abelev:2012sea}. 
\end{itemize}
\begin{table}
\centering
\begin{tabular}{|l|c|llll|} 
\hline
Expt                   & $\surd s$
                       & $\sigma_{tot} \text{ [mb]}$
		       & $\sigma_{el} \text{ [mb]}$
		       & $\sigma_{in} \text{ [mb]}$
		       & $\sigma_{sd} \text{ [mb]}$  \\ 
\hline
\multirow{3}{*}{UA5}   & 200 GeV
                       &
		       &
		       &
		       & $4.8 \pm 0.9$               \\
                       & 546 GeV
		       &
		       &
		       &
		       & $5.4 \pm 1.1$               \\
                       & 900 GeV
		       &
		       &
		       & $50.3 \pm 1.1$
		       & $7.8 \pm 1.2$               \\
\hline
\multirow{2}{*}{E710}  & 1.02 TeV
                       & $61.1 \pm 9.9$
		       &
		       &
		       &                             \\
                       & 1.8 TeV
		       & $78.3 \pm 5.9$
		       & $19.6 \pm 3.0$
		       &
		       & $8.1 \pm 1.7$               \\ 
\hline
\multirow{4}{*}{ATLAS} & 7 TeV
                       & $95.4 \pm 1.4$
		       & $24.0 \pm 0.6$
		       &
		       &                             \\
                       & 7 TeV
		       &
		       &
		       & $69.4 \pm 7.3$
		       &                             \\
                       & 8 TeV
		       & $96.1 \pm 0.9$
		       & $24.3 \pm 0.4$
		       &
		       &                             \\
                       & 13 TeV
		       &
		       &
		       & $78.0 \pm 3.0$
		       &                             \\ 
\hline
\multirow{2}{*}{ALICE} & 2.76 TeV
                       &
		       &
		       & $62.8 \pm 3.4$
		       & $12.2 \pm 4.6$              \\
                       & 7 TeV
		       &
		       &
		       & $73.2 \pm 4.3$
		       & $14.9 \pm 4.7$              \\ 
\hline
\multirow{2}{*}{LHCb}  & 7 TeV
                       &
		       &
		       & $68.7 \pm 4.9$
		       &                             \\
                       & 13 TeV
		       &
		       &
		       & $75.4 \pm 5.4$
		       &                             \\ 
\hline
\multirow{2}{*}{TOTEM} & 7 TeV
                       &
		       &
		       & $73.7 \pm 3.4$
		       &                             \\
                       & 13 TeV
		       & $110.6 \pm 3.4$
		       & $31.0 \pm 1.7$
		       &
		       &                             \\ 
\hline
\end{tabular}
\caption{High energy $p\porpb$ experimental data set supplemented by data available in~\cite{Tanabashi:2018oca} }
\label{tab:dataset}
\end{table}

A few caveats about our data selection are in order.
We use measured data from experiments that quote both statistical and systematic
errors, and combine them in quadrature.
We omit $pp$ cross-section measurements from cosmic-ray experiments because
the reconstruction of these events uses Monte Carlo showering codes such as
\texttt{SIBYLL} \cite{Fedynitch:2018cbl} and \texttt{QGSJET-II}
\cite{Ostapchenko:2010vb} which use the eikonal unitarisation scheme.

As discussed in \cite{Bhattacharya:2020lac}, there is considerable tension
amongst the total and elastic cross sections at the same or similar energies from different
experiments (see also \cite{Cudell:1996sh,Block:2005qm}).
We quantify these inconsistencies by fitting each kind of cross
section with a quadratic polynomial in $\log s$, the resulting
\chisq\ shown in Table \ref{tab:data1}.

We particularly note that at centre-of-mass energies of 7 and 8 TeV, total
and elastic cross sections from TOTEM are consistently $1\sigma$ higher
than those from ATLAS.
The low statistics we have to work with prevents us from determining which
experimental results are the outliers, so we shall continue to use all of the
data points with the cognisance that the resulting \chisq\ will inevitably be
high.
When including single diffractive data, this enforces a baseline minimum of
$\chi^2 = 49.6 $ for 43 data points.
\begin{table}[t]
  \centering
  \begin{tabular}{|c|c|c|}
    \hline
    dataset &number of points &$\chi^2$\\
    \hline
    $\sigma_{tot}$&18  & 21.7\\
    $\sigma_{el}$&11  & 21.3\\
    $\sigma_{in}$&8  & 4.1\\
    $\sigma_{sd}$&6  & 2.6\\
    \hline
  \end{tabular}
  \caption{\label{tab:data1} The values of $\chi^2$ resulting from independent fits to quadratic polynomials
  in $\log(s)$, illustrating the tensions in some parts of the dataset.         }
\end{table}

Furthermore, we do not include double diffractive cross-section measurements \cite{Ansorge:1986xq,Affolder:2001vx,Abelev:2012sea} in our fits
since a proper description of these cross sections has so far eluded any theoretical description.
We have checked that our models are not able to reproduce these, even if we free all possible parameters.
We show the discrepancy in Fig.~\ref{fig:bfxsec}.

\section{\label{sec:result}Results}

\begin{figure}[t]
\centering
\subfloat{\includegraphics[width=0.5\textwidth]{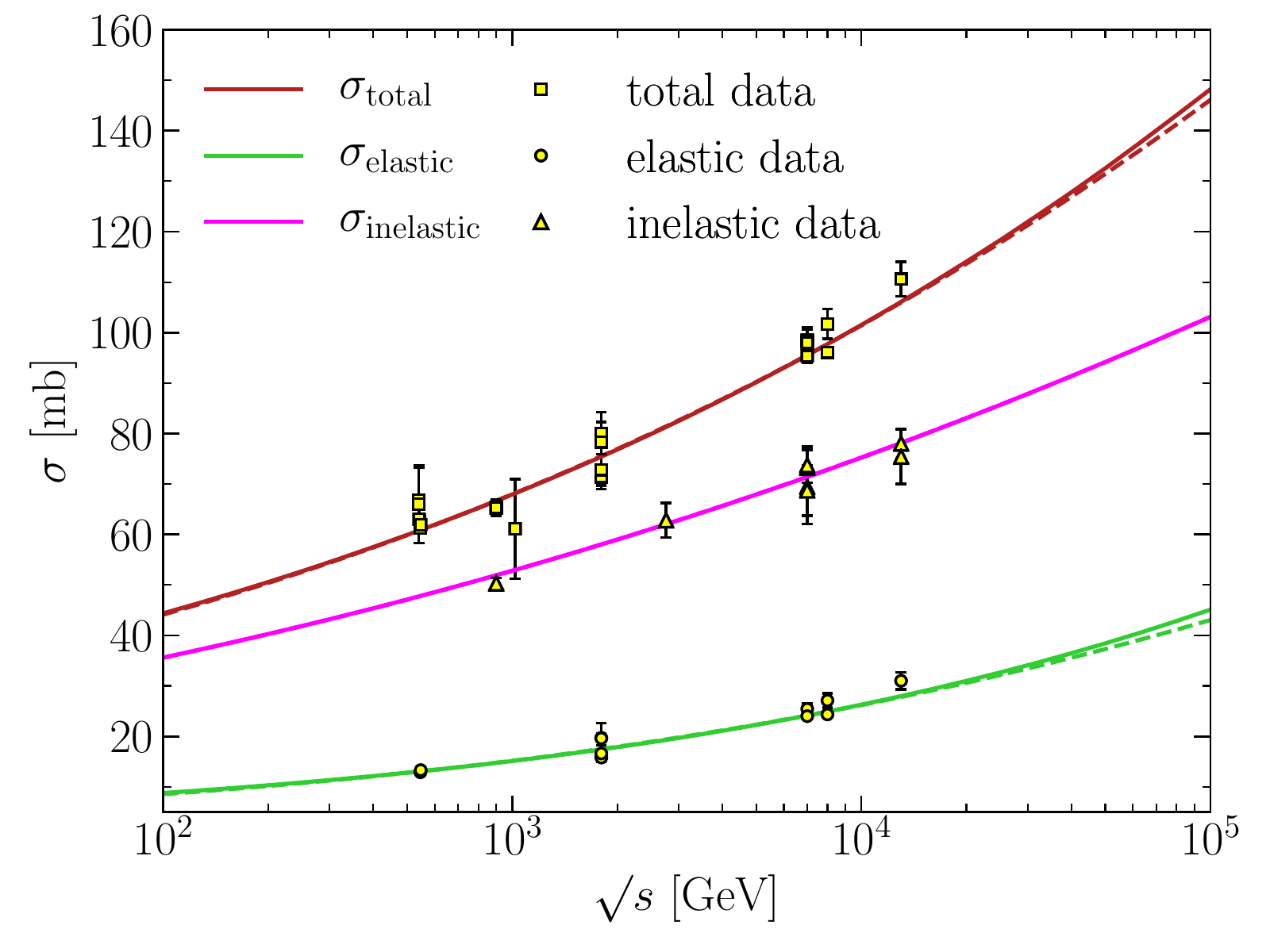}}
\subfloat{\includegraphics[width=0.5\textwidth]{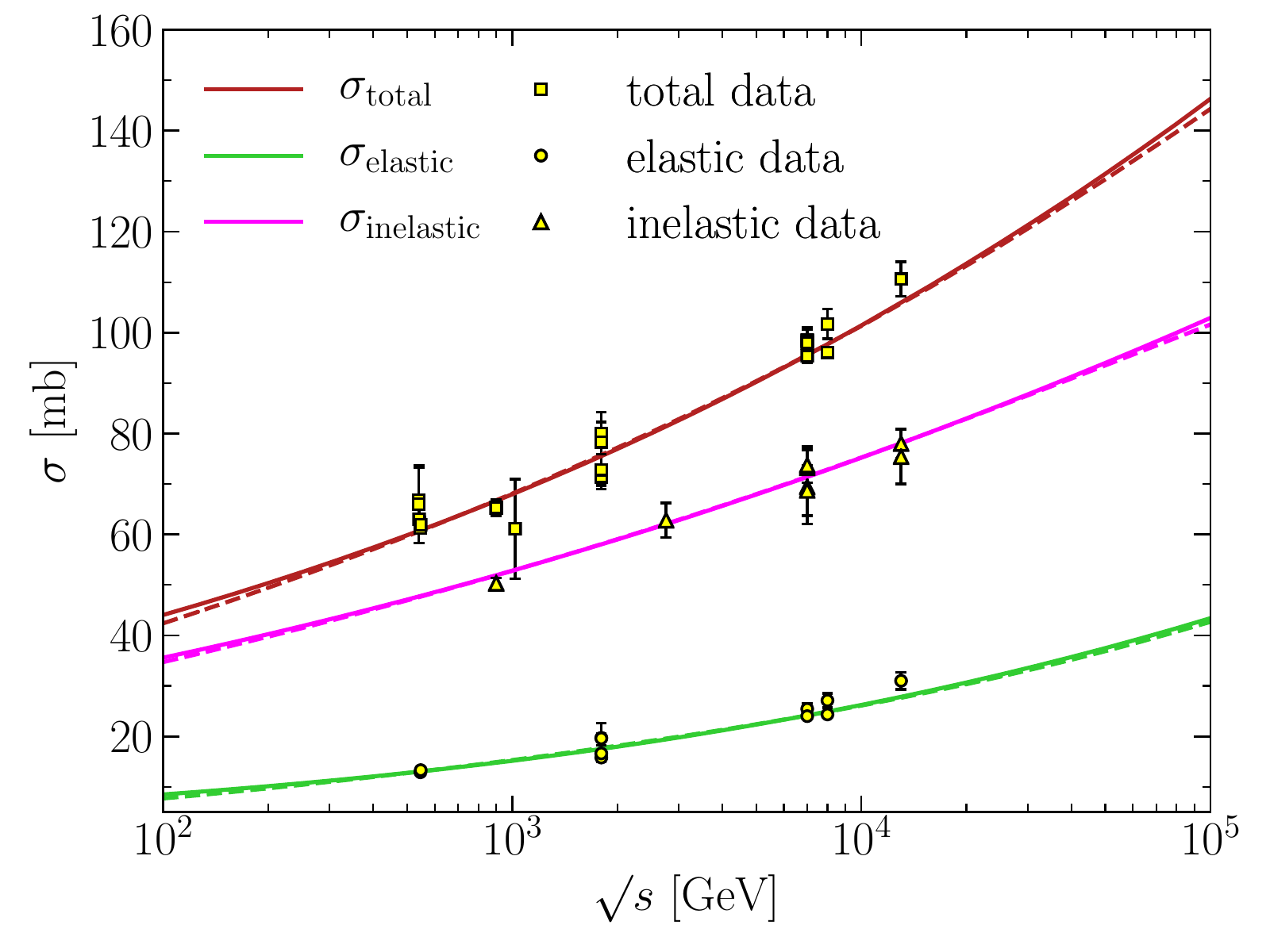}} \\
\subfloat{\includegraphics[width=0.5\textwidth]{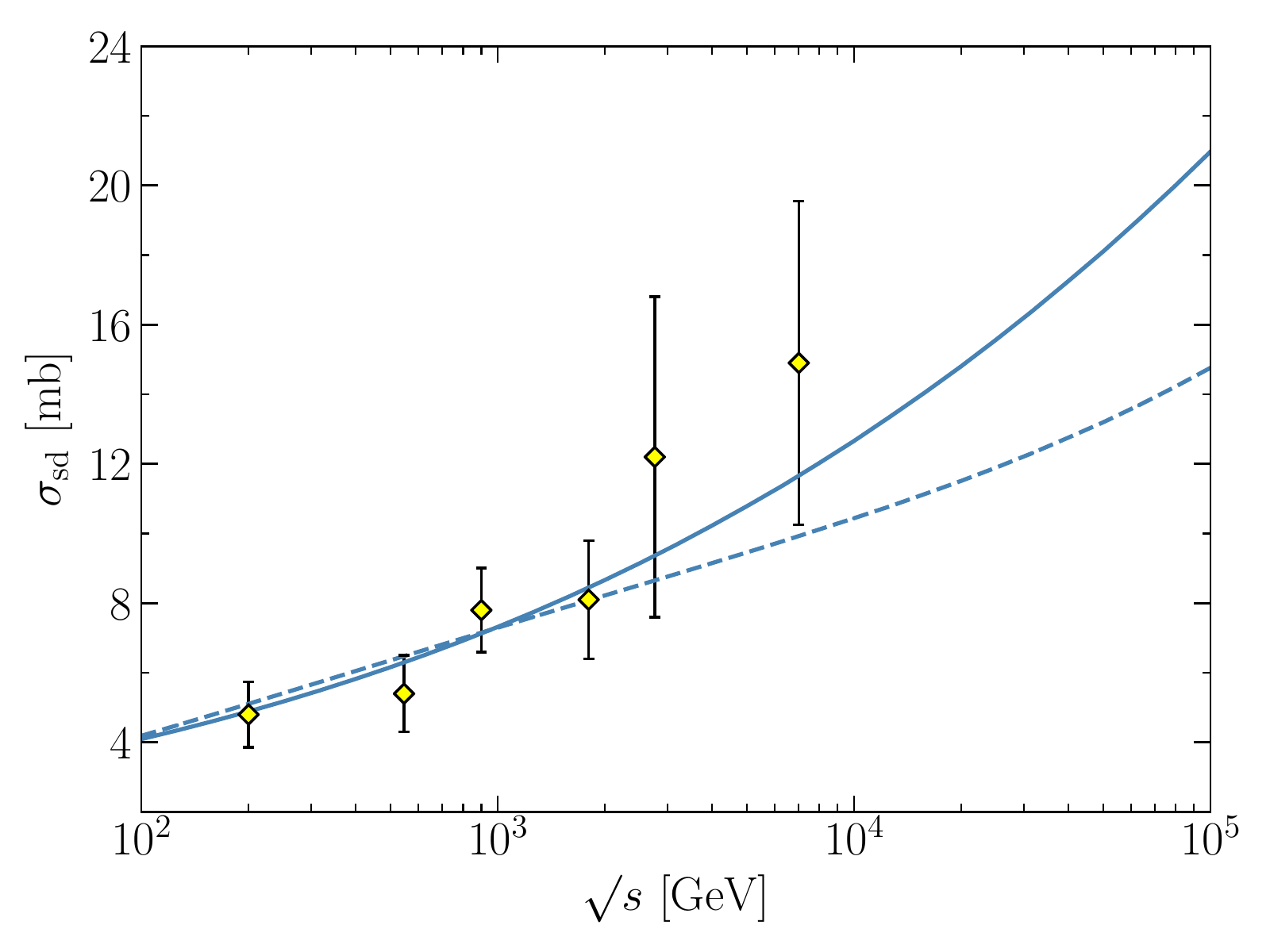}}
\subfloat{\includegraphics[width=0.5\textwidth]{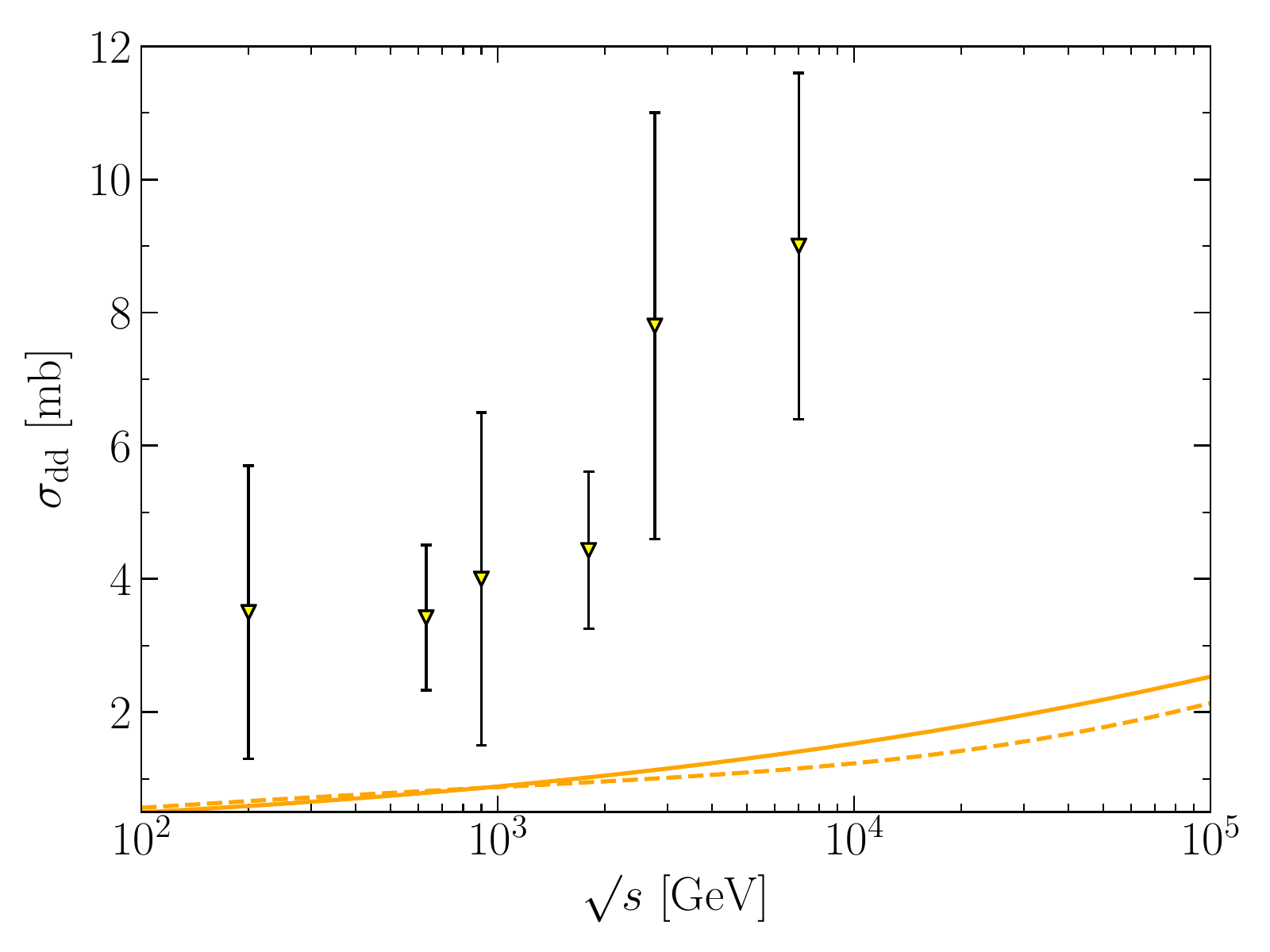}}
\caption{\label{fig:bfxsec}Top left: Total, elastic, and inelastic
         cross sections obtained with best-fit
         parameters for the U-matrix (solid curves) and the eikonal schemes
         (dashed curves) without using single-diffractive data.
         Top-right: Same as top-left but when single diffractive data is
         included in the fits.
         Bottom-left: Single diffractive cross-sections for best-fit values of
         the parameters when using the U-matrix (solid curves) and
         eikonal schemes (dashed curve).
         Bottom-right: Double diffractive cross-sections, which are not well
         fit by either scheme.}
\end{figure}
We give the results of our fits in Fig.~\ref{fig:bfxsec} and Table~\ref{tab:fitsd}. 
We obtain equivalent fits for the U matrix and the eikonal, with respective values of the $ \chisq/\text{d.o.f}$
of 1.316 and 1.328. As discussed in Sec. \ref{sec:fits}, these high values are driven by disagreements in the elastic data at the high energies.
\begin{table}[h]
  \footnotesize
  \begin{center}
\begin{tabular}{|c||c|c|c|c|c|c|}
   \hline
  Scheme   & $\epsilon $
            & $\apr $ (GeV$^{-2}$)
            & $ g_{DD} $
            & $ t_{DD} $ (GeV$^2$)
            & $ \theta $ (rad)
            & $ \chisq/\dof $\\
            \hline
    U-matrix & $ 0.11 \pm 0.08 $
            & $ 0.35 \pm 0.05 $
            & $ 6.3 \pm 1.3  $
            & $ 2.2 \pm 0.4 $
            & $ 0.11 \pm 0.02$
            & $ 1.316 $ \\
   Eikonal  & $ 0.12 \pm 0.04 $
            & $ 0.31 \pm 0.10 $
            & $ 8.81 \pm 0.12 $
            & $ 1.37 \pm 0.05$
            & $ 0.20 \pm 0.02$
            & $ 1.328 $ \\
   \hline
\end{tabular}

  \end{center}
  \caption{\label{tab:fitsd}\chisq/\dof\ and best-fit parameters obtained using the
           eikonal ($ \omega = 1 $) and U-matrix ($ \omega^\prime = 1/2 $)
           unitarisation schemes with single diffractive
           data.
           The parameters of the $pp$ vertex are fixed to the central values of
           Table \ref{tab:el}. 
          }
\end{table}
With this understanding, it is clear that the data allows for
U-matrix unitarisation scheme.
Either scheme describes the total and elastic cross sections equally well;
however, the U-matrix scheme provides a slightly better fit to the
high-energy single diffractive data than does the eikonal, as can be seen in 
Fig.~\ref{fig:bfxsec}.

The parameters of the pomeron trajectory are not affected by the inclusion of the
diffractive data, as they have a much lower weight than the elastic data. The parameters
linked to the diffractive state are consistent with the physical picture underlying our model:
the diffractive state is slightly bigger than the proton hence its scale $t_{DD}$ is slightly lower
than $t_{pp}$.

As noted previously, the double diffractive cross sections $ p\porpb \to 2\,X $ \cite{Ansorge:1986xq,Affolder:2001vx,Abelev:2012sea} are not fitted well by either of the unitarisation schemes.
We show this in Fig.~\ref{fig:bfxsec} (bottom-right panel).

Multiple experiments have investigated the ratio of the real part of the elastic
scattering amplitude to its imaginary part at different centre-of-mass energies.
Although we do not use $\rho$ data in our fits, we can predict its
values at different $ \sqrt{s} $ using our best-fit parameters and compare these
predictions against the experimental data.
We find that the values of $\rho$ and its slowly-falling shape as a function of
$ \sqrt{s} $ are largely consistent with experimental data
between 100 GeV and 7 TeV (see \eg\ \cite{Tanabashi:2018oca}).
We predict $ \rho = 0.131 \pm 0.009 $ for either unitarisation scheme at $ \surd s = 13 $ TeV.
This agrees with the result $ \rho = 0.14 $ in \cite{Donnachie:2019ciz};
however, it is in tension with the value of $\rho \approx 0.10 $ obtained for the
13 TeV TOTEM data both by the collaboration itself \cite{Antchev:2017yns}
and in \cite{Cudell:2019mbe}.

\begin{figure}[htb]
\centering
\includegraphics[width=0.7\textwidth]{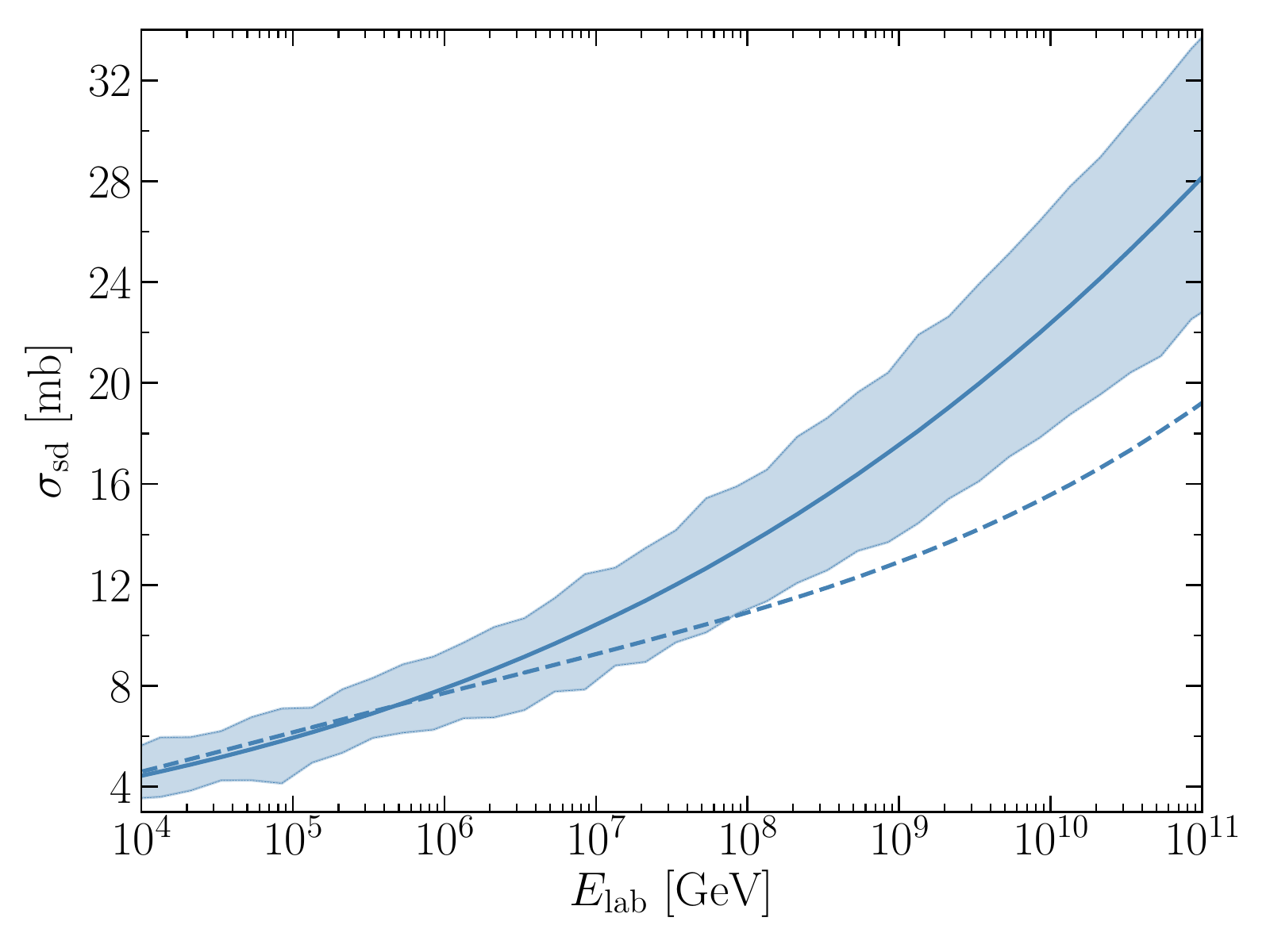}
\caption{\label{fig:uhecosmic} The growth of the single-diffractive cross
         section with \emph{lab} energies up to $\sqrt{s}  = 10^{11}$ GeV for
         both the U-matrix unitarisation scheme (solid curve) and the eikonal (dashed
         curve).
         We show a $1\sigma$ error band in the U-matrix case. The corresponding band is
         similar in the eikonal case, but we omit it for clarity.}
\end{figure}

Despite their equivalence for existing data, the two schemes give significantly different
predictions for the single-diffractive cross section at ultra-high energies. Unlike the total, elastic, and inelastic cross sections, the single diffractive
cross section obtained using the eikonal scheme
is noticeably different from that obtained using the U-matrix, with the
former exhibiting a slower growth with energies than the latter, as shown in
Fig.~\ref{fig:uhecosmic}.
This difference is especially significant for ongoing cosmic-ray experiments
measuring the $pp$ cross-section at high energies from tens of TeV up to the
GZK cut-off, $ E_\text{lab} \approx 5 \times 10^{10} $ GeV. As the single-diffraction
 is the parent process to forward pions, and hence to forward muons, it seems that considering
 different unitarisation schemes would lead to different muon multiplicities at ultra-high energies.

\section{Conclusions}
We have shown how the scheme proposed by Gotsman, Levin and 
Maor \cite{Gotsman:1999xq,Gotsman:2014pwa} could be adapted to other unitarisation schemes.
We have also shown how the vertices of the mixed states could be deduced from those of the
proton, allowing a more constrained parameterisation.

Using up-to-date collider data on $p\porpb$ total, elastic, and single
diffractive cross sections, including 13 TeV data from recent LHC experiments,
we have determined best fits to the parameters governing
these cross sections in the context of different unitarisation schemes.
Specifically, we have shown that the U-matrix scheme fits the data as well as
the more ubiquitous eikonal scheme.
In fact, the fits have a slight preference for the U Matrix. This difference is driven 
by  the single diffractive
cross section, especially at high energies, while the best-fit total and elastic
cross sections are nearly identical up to energies of 13 TeV when using either
of these schemes.

A consequence of the indifference of the elastic cross section to the choice of
the unitarisation scheme up to tens of TeV is that values of the $ \rho $
parameter remain largely unaffected by the choice of the scheme too.
We use our best-fit parameters to compute this parameter across different
energies, and find that the corresponding values conform to existing data, to the
exception of the TOTEM measurement at 13 TeV.

We have also analysed how the fits improve if one uses the generalised eikonal
and U-matrix schemes and we find that these generalisations --- at the cost of
an additional free parameter ($ \omega \text{ or } \omega^\prime $) --- do not
improve the fits significantly.

The upshot of our analysis is that the overall best-fit cross section, in light
of up-to-date collider data, is obtained using amplitudes unitarised via the
U-matrix scheme.
The resulting $ p\porpb $ single diffractive cross section shows a sharper
growth at high
energies than does the one obtained using the more commonly used eikonal scheme,
and unitarisation could have an impact on the description of ultra-high-energy cosmic-ray 
showers.

\acknowledgments{
AB is supported by the  Fonds  de  la  Recherche  Scientifique-FNRS,  Belgium,
under grant No.~4.4503.19.
AB is thankful to the computational resource provided by Consortium des
Équipements de Calcul Intensif (CÉCI), funded by the Fonds de la Recherche
Scientifique de Belgique (F.R.S.-FNRS) under Grant No.~2.5020.11 where a part of
the computational work was carried out. AV is supported by U.S. DOE Early Career Research Program under FWP100331.
}

\bibliographystyle{JHEP}
\bibliography{refs}
\end{document}